\documentclass{article}

\usepackage{adjustbox}

\usepackage{arxiv}
\usepackage{amsmath}
\usepackage[utf8]{inputenc} 
\usepackage[T1]{fontenc}    
\usepackage{hyperref}       
\usepackage{url}            
\usepackage{booktabs}       
\usepackage{amsfonts}       
\usepackage{nicefrac}       
\usepackage{microtype}      
\usepackage{lipsum}		
\usepackage{graphicx}
\usepackage{natbib}
\usepackage{doi}
\usepackage{comment}
\usepackage{siunitx}

\title{Towards a general description of the cavitation threshold in acoustic systems}

\author{ \href{https://orcid.org/0000-0002-1219-3263}{\includegraphics[scale=0.06]{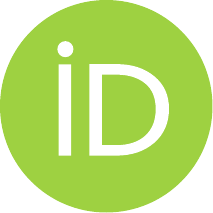}\hspace{1mm}Gianmaria ~Viciconte}\\
        Clean Combustion Research Center\\
	Department of Physical Science and Engineering\\
	King Abdullah University of Science and Technology\\
	Thuwal 23955, Saudi Arabia \\
	\texttt{gianmaria.viciconte@kaust.edu.sa} \\
 \And
 \href{https://orcid.org/0000-0002-9805-9291}{\includegraphics[scale=0.06]{orcid.pdf}\hspace{1mm}Paolo ~Guida} \\
        Clean Combustion Research Center\\
	Department of Physical Science and Engineering\\
	King Abdullah University of Science and Technology\\
	Thuwal 23955, Saudi Arabia \\
	\texttt{paolo.guida@kaust.edu.sa} \\
\And
 \href{https://orcid.org/0000-0003-1613-6052}{\includegraphics[scale=0.06]{orcid.pdf}\hspace{1mm}Tadd ~Truscott} \\
        Splash Lab\\
	Department of Physical Science and Engineering\\
	King Abdullah University of Science and Technology\\
	Thuwal 23955, Saudi Arabia \\
	\texttt{tadd.truscott@kaust.edu.sa} \\
 \And
 \href{https://orcid.org/0000-0003-1999-2831}{\includegraphics[scale=0.06]{orcid.pdf}\hspace{1mm}William L. ~Roberts} \\
        Clean Combustion Research Center\\
	Department of Physical Science and Engineering\\
	King Abdullah University of Science and Technology\\
	Thuwal 23955, Saudi Arabia \\
	\texttt{william.roberts@kaust.edu.sa} \\ 
}

\hypersetup{
pdftitle={Towards a general description of the cavitation threshold in ultrasonic systems},
pdfsubject={q-bio.NC, q-bio.QM},
pdfauthor={Gianmaria Viciconte, Paolo Guida, Tadd Truscott, William L. Roberts},
pdfkeywords={Ultrasonicallly-Induced Cavitation, Cavitation Threshold, Acoustic Analogy },
}

\begin{document}

\maketitle

\begin{abstract}
Traditionally, cavitation has been related to the ratio between flow velocity and pressure gradient in the case of hydrodynamic cavitation, or some combination of vapor pressure and surface tension. However, both formulations present a large discrepancy with experimental data for cases where acoustic waves induce cavitation. The present study aims to identify a more suitable cavitation threshold for such cases. The methodology adopted in this work consists of a combination of visualization with high-speed cameras and direct measurements using a hydrophone. The data collected confirmed that vapor pressure is not a proper indicator of cavitation occurrence for an acoustic system characterized by high frequencies.  The main reason behind the inability of vapor pressure to predict incipient cavitation in acoustic systems is that they evolve very quickly toward strong gradients in pressure, and the quasi-static assumptions used by traditional models are not valid. Instead, the system evolves towards a metastable state \citep{Brennen1995}, where the liquid exhibits an elastic behavior and can withstand negative pressures.  A new cavitation number was defined that accounts for the tensile strength of the liquid. However, to provide a complete description of the cavitation threshold, future experiments, in a wide range of acoustic frequencies, are still necessary. An acoustic analogy is also proposed for the description, with the same framework, of an impulsive cavitation phenomenon. 

\end{abstract}


Cavitation induced by acoustic waves is a physical phenomenon used, nowadays, in a wide range of industrial processes and laboratory applications, from the pharmaceutical and clinical sector to biomass treatment and heavy oil upgrading \citep{guida2022numerical, sharma2003effect, flores2021ultrasound}. Most applications are based on high-power ultrasound transducers to generate cavitation. In this type of system, vapor-filled cavities form due to the propagation of acoustic waves in a liquid domain \citep{suslick1999acoustic, kozmus2022characterization, birkin2020cavitation}. Once formed, the vapor cavities oscillate, and collapse, depositing the energy stored in confined regions of space. The confined energy release induces the formation of hot spots, characterized by high pressure and temperature \citep{Suslick2008}, and the production of radicals species (Sonochemistry) \citep{suslick1990sonochemistry,didenko1999hot}. The collapse of the vapor cavities also induces peculiar fluid dynamic phenomena, like shock waves and micro-jets \citep{wagterveld2011visualization}. These phenomena can potentially increase the reaction rate of various chemical processes, enhancing the mixing in a multi-phase system, the fragmentation in a solution, or crystallization \citep{birkin2022probing}.

Laboratory and industrial processes, based on acoustic cavitation, are commonly carried out using a variety of ultrasound reactor designs, working frequencies, and operative conditions. However, the optimal combination of working parameters for a given process is unclear. To improve these systems' design and efficiency, it is crucial to understand the physical mechanism behind cavitation onset \citep{smirnov2022analysis}.
Traditionally, the cavitation threshold is established by considering that the phase transition occurs when pressure reaches the saturation pressure of the liquid medium. The most widely adopted Cavitation number ($Ca$) is defined similarly to the Euler number:
\begin{equation}
    Ca=\frac{p_r-p_v}{\frac{1}{2}\rho v^2},
\label{equation_0_1}
\end{equation}
where $p_r$ is the reference pressure, $p_v$ is the vapor pressure of the liquid, $\rho$  is the liquid density, and $v$ is the local velocity of the flow. This number is used to characterize hydrodynamic cavitation phenomena \citep{brennen2011hydrodynamics,arndt1981cavitation}. Recently, an alternative cavitation number, based on the vapor pressure as a cavitation threshold, has been introduced by \citet{fatjo2016new} and \citet{Pan2017}. In this dimensionless number, the acceleration ($a$) and the depth of the liquid column ($h$) appear as scaling factors:
\begin{equation}
    Ca=\frac{p_r-p_v}{\rho a h}.
\label{equation_0_2}
\end{equation}
This number can successfully predict cavitation phenomena in systems set into motion by an impulsive force \citep{Pan2017}. 
While a cavitation number, based on the saturation vapor pressure, can describe certain phenomena effectively, it does not predict the onset of cavitation caused by the propagation of acoustic waves in a liquid medium. When an acoustic wave travels through a medium, it generates compression and rarefaction regions \citep{lurton}. If high frequencies characterize the waves, the thermodynamic system evolves quickly, with strong gradients of the thermodynamic variables. Such a system is far from the theoretical hypothesis on which the definition of vapor pressure is based. Indeed, the theoretical framework and the experimental data used to build the phase diagrams, in the context of classical thermodynamics, are based on the assumption of a quasi-static process \citep{Gyf}. A quasi-static process evolves sufficiently slowly to guarantee that the system remains in thermodynamic equilibrium at every instant of time \citep{woods1959foundations,moebs2016university}.

In a thermodynamic system where the quasi-static assumption is invalid, an isothermal depressurization may lead to a metastable state in which the liquid withstands pressures lower than the saturation pressure \citep{Brennen1995}, without turning into vapor. The pressure potentially reaches negative values with the liquid being in a state of tension \citep{Brennen1995}. Berthelot \citep{berthelot1850quelques} first attained this condition in a controlled environment, in his work from 1850, demonstrating that purified water could withstand a tension of 50 bar before the “rupture”.
\newline

Brennen \citep{Brennen1995} introduced an elastic analogy stating that liquids can behave like elastic media since they can withstand tension. The elastic property of liquids is described by the Deborah number ($De$), which is the ratio between the characteristic time for the self-diffusion of a molecule and the characteristic time of the applied force \citep{Pelton2013, noirez2012identification, urick}. It is possible to define the Tensile Strength of a liquid as the maximum Tensile Stress (tension) that a liquid medium can withstand, while being ”stretched”, before changing phase to vapor. This theory supports the fact that the tensile strength, instead of the vapor pressure, should be considered as a cavitation threshold for the phenomena where the nucleation is induced by acoustic waves, especially at high frequencies, where the liquid increases its tendency to behave as an elastic medium. Furthermore, this leads us to consider that the acoustic cavitation threshold should depend on the frequency of the acoustic wave. 

\begin{figure}[h!]
    \centering
    \includegraphics[width=0.60\linewidth]{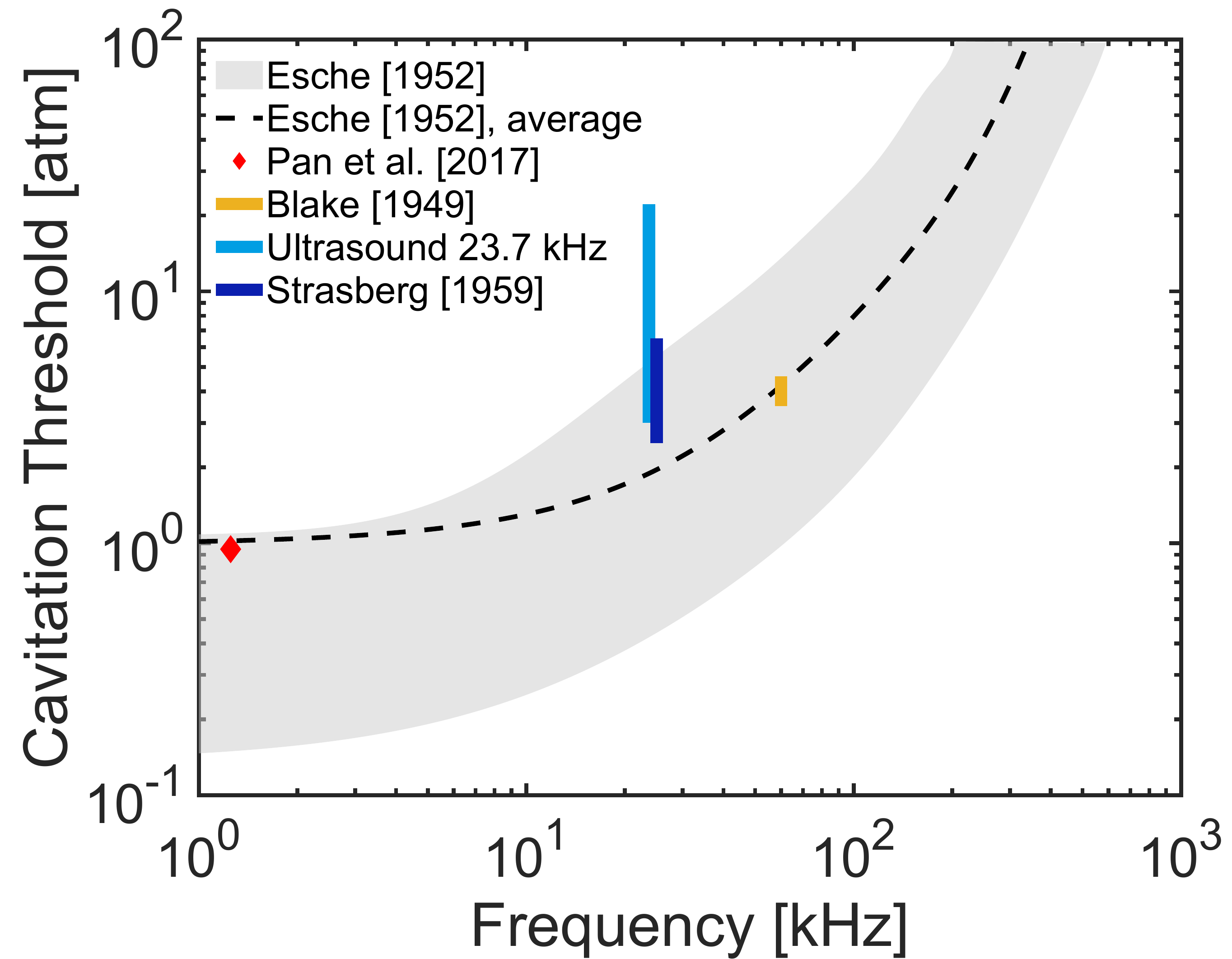}
    \caption{Experimental chart of the acoustic cavitation threshold of water as a function of the frequency. The data are related to measurements done by several authors \citep{urick,blake1949onset,strasberg1959onset,esche1952untersuchung}.} 
    \label{figure_1}
\end{figure}

In this regard, \citet{urick} proposed a direct dependency on the frequency of the sound wave (Fig. \ref{figure_1}). This dependency predicts a significant increase in the cavitation threshold of water as the frequency increases. The occurrence of two mechanisms justifies the trend observed by \citet{urick}. The first one is the tendency of the liquid to increase its elastic response at a higher forcing frequency (high $De$ number). The second relates to the characteristic time of growth of a macroscopic bubble, starting from a nucleus. The chart, proposed by Urick (Fig. \ref{figure_1}), is based on the experimental data of several authors. \citet{esche1952untersuchung} measured the cavitation threshold of water at various frequencies. The shaded area, on the chart in Fig. \ref{figure_1}, represents the range of observed values, while the dashed curve is the estimated average \citep{esche1952untersuchung}. \citet{strasberg1959onset}, using a forcing frequency of 25 kHz, detected a threshold ranging from 2.5 atm in tap water saturated with air to 6.5 atm in degassed water (blue bar in Fig. \ref{figure_1}). \citet{blake1949onset} measured values ranging from 3.5 atm to 4.6 atm for air saturated and degassed water (orange bar in Fig. \ref{figure_1}), using a frequency of 60 kHz. These experimental values are an order of magnitude lower than those obtained in controlled environments, using highly purified liquids \citep{berthelot1850quelques, dixon1909note}. Recently, \citet{caupin2012exploring} obtained a cavitation pressure of -20 MPa. However, these values, despite the controlled environment and the purified liquids, are far away from the theoretical limit established by the Homogeneous Nucleation Theory \citep{Brennen1995}, which predicts the rupture of pure water at values in the order of magnitude of \(10^4\) atm. This discrepancy suggests that contaminants play a crucial role by decreasing the nucleation energy barrier, in every laboratory system and, in general, in real media \citep{Brennen1995}. 
\newline

In an attempt to unify the theoretical understanding behind the onset of acoustic cavitation, we studied a system operating at 24 kHz. A rigorous and repeatable methodology for defining the onset of cavitation is defined from visual and acoustic data. Furthermore, based on the experimental observation and on the Rayleigh analytical model \citep{rayleigh1896theory}, a Cavitation number is proposed \citep{viciconte2023unifying}. This dimensionless number contains the tensile strength as a cavitation threshold and the frequency, of the acoustic wave, as a scaling factor. Afterward, through an acoustic analogy, the same framework is applied to a case of cavitation induced by impulsive motion \citep{Pan2017}.

\section{Experimental procedure}
\begin{figure}[h!]
\centering
\includegraphics[width=0.65\linewidth]{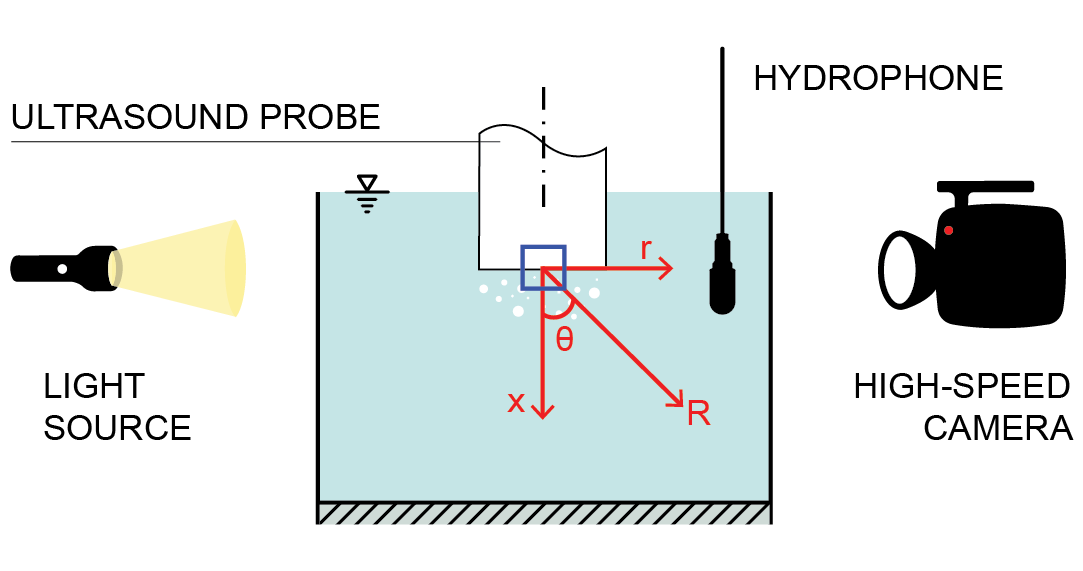}
\caption{Schematic illustration of the backlighting setup used for the visualization experiments. The blue square indicates the high-speed camera field of view.}
\label{Figure_end} 
\end{figure}

Visualization experiments have been conducted using a backlighting technique. A schematic illustration of the experimental setup is visible in Fig. \ref{Figure_end}. 
A Hielscher device UP400S (\SI{24}{kHz}) was used to generate ultrasound waves in the liquid domain. Additional details can be found in the section \textit{Methodology}. 

\begin{figure}[h!]
\centering
\includegraphics[width=0.75\linewidth]{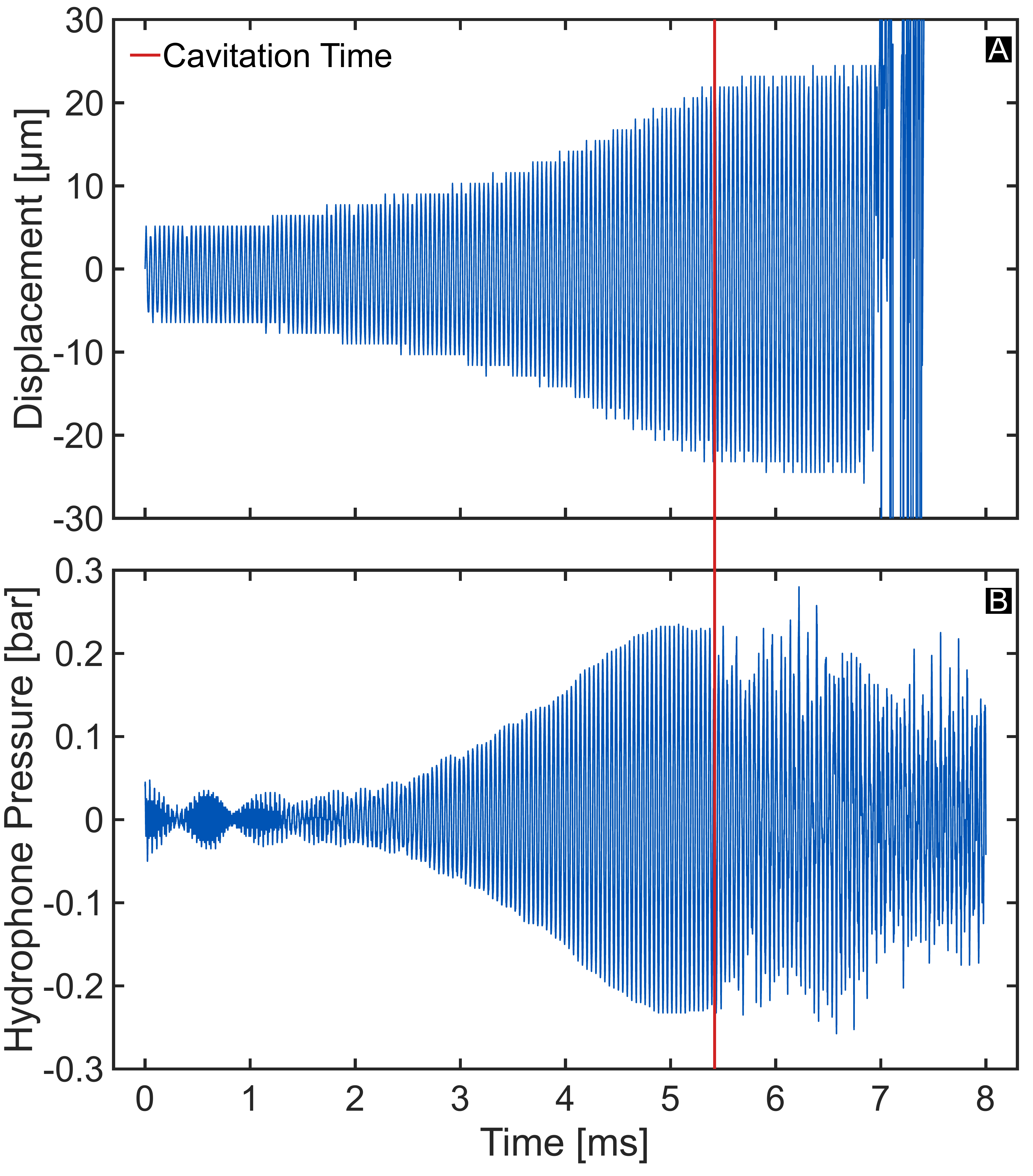}
\caption{(A) Result of the tracking procedure on one experimental observation (nominal amplitude set on the device: 30 \%). The blue function represents the displacement of the probe over time, while the vertical red line represents the time at which cavitation happens. (B) The hydrophone measured acoustic pressure (nominal amplitude set on the device: 30 \%). The vertical red line indicates the cavitation time defined with the visual tracking procedure.} 
\label{Figure3_1} 
\end{figure}

The displacement amplitude ($A$) of the ultrasound probe undergoes a transient state where it gradually increases over time, until reaching a steady state (Fig. \ref{Figure3_1} (A)). The high-speed camera was triggered, by the intensity of the hydrophone acoustic signal near cavitation inception. The probe's displacement must be precisely measured to measure the pressure at which cavitation onset happens. Since the displacement is in the order of magnitude of micrometers, an objective lens (10X magnification), providing a resolution of \SI{1.98}{\micro m}/pixel, was used.
The sequence of images in Fig. \ref{fig0} (A, B) shows the portion of the domain framed by the high-speed camera (blue outlined area in Fig. \ref{Figure_end}). The movement of the probe tip and the nucleation of the first vapor cavity are visible. It is important to note that the nucleation of the first cavity always happens during the rising motion of the ultrasound probe. This is because of the refraction of the liquid medium induced by the probe rising, with the consequent formation of a low-pressure region. 
The cavitation onset, on the surface of the probe tip, is better observable by framing a wider region of the domain (Fig. \ref{fig0} (C, D)). This has been done by using a Nikon MICRO Lens, having a focal length of \SI{105}{mm}. 

\begin{figure}[h!]
\centering
\includegraphics[width=0.75\linewidth]{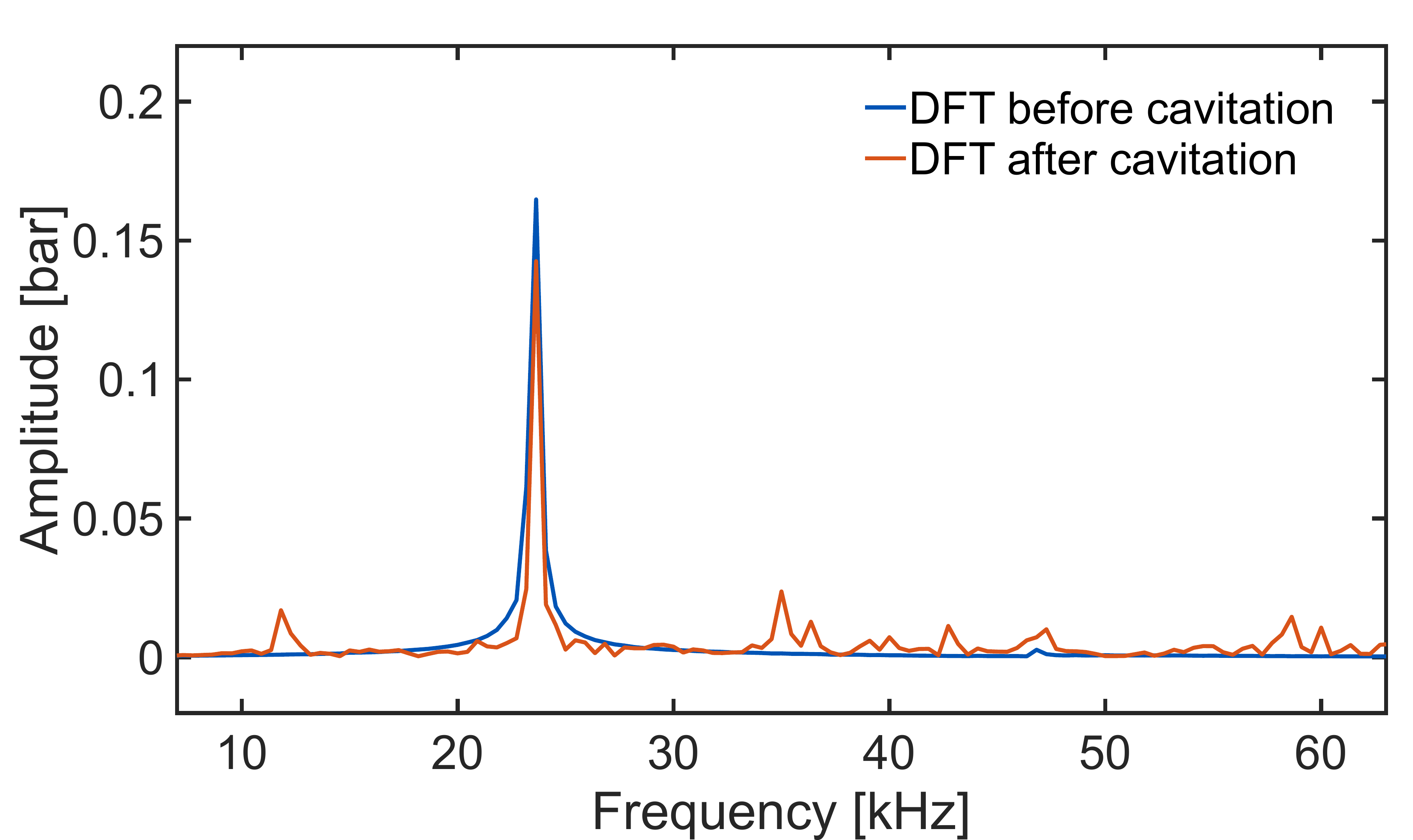}
\caption{DFT frequency spectrum of the acoustic signal measured by the hydrophone. The frequency spectrum, related to the signal before cavitation happens (blue line), shows peaks just at the vibration frequency of the ultrasound probe and at its multiples. Instead, the spectrum after cavitation (orange line) has additional peaks and broad-band noise due to the presence of cavitation bubbles in the liquid domain.}
\label{Figure_fourier} 
\end{figure}

\begin{figure*}[t!]
    \centering
    \includegraphics[scale=0.95]{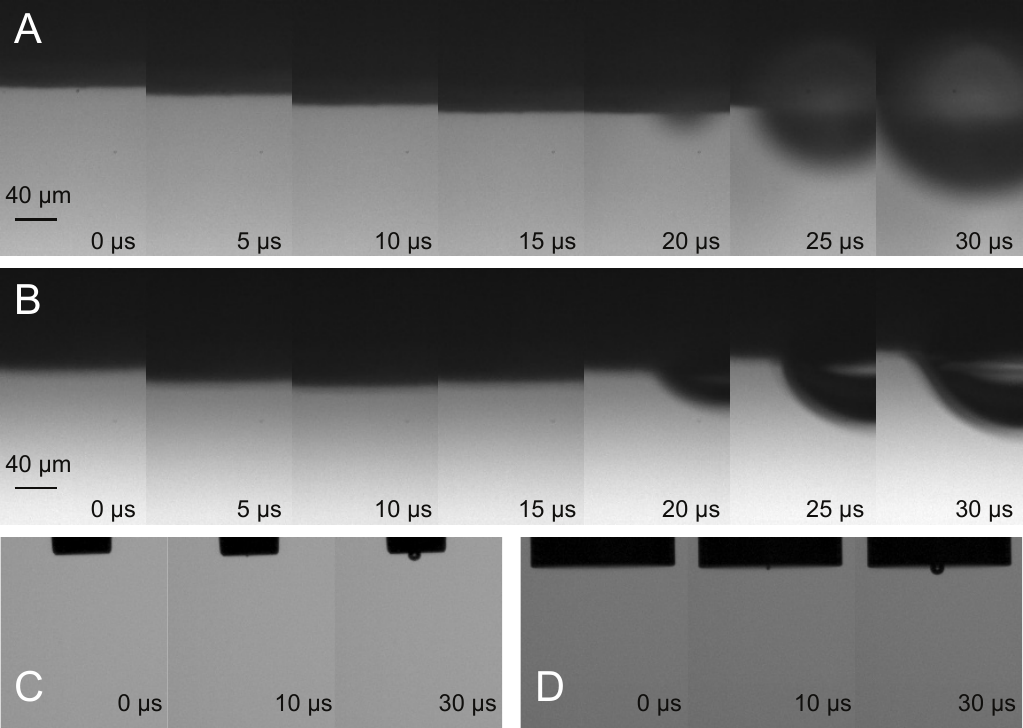}
    \caption{(A) Sequence of frames recorded by using the high-speed camera, coupled with a 10X objective lens. It is clearly visible the movement of the probe tip and the nucleation of the first cavity attached to the tip surface. This sequence is related to a 3 mm probe with a device nominal amplitude of 70 \%. (B) Sequence of frames related to a 3 mm probe with a nominal amplitude of 100 \%. (C) Sequence of images framed by using a Nikon MICRO Lens (105 mm). The nucleation of a cavity, attached to the surface, on the axis of a cylindrical probe (3 mm diameter) was captured. (D) Nucleation of a bubble on the tip surface of a cylindrical probe, having a diameter of 7 mm (recorded by using a 105 mm Nikon MICRO Lens).}
    \label{fig0}
\end{figure*}

\begin{figure}[h!]
\centering
\includegraphics[width=0.75\linewidth]{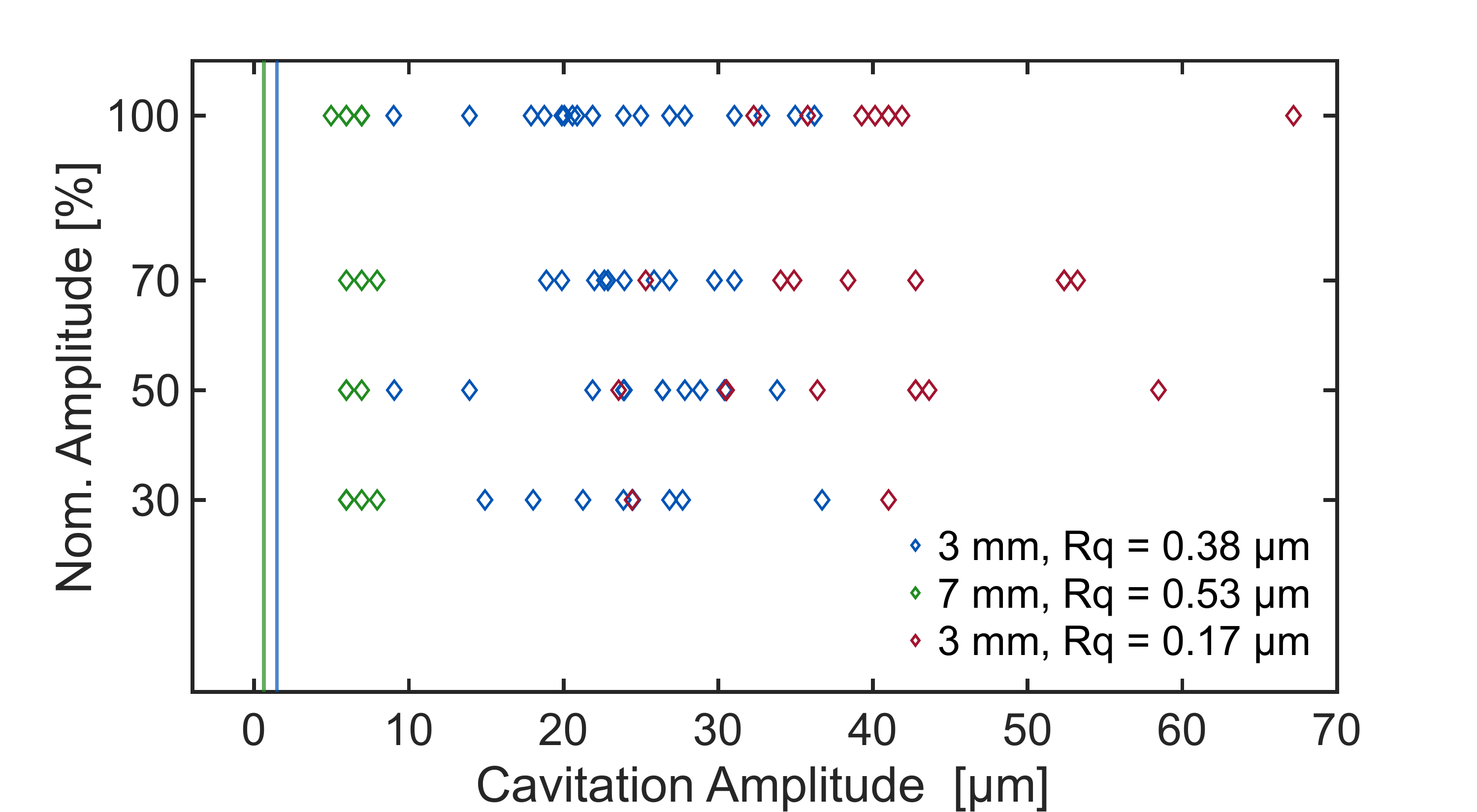}
\caption{Scatter plot representing the displacement amplitude of the probe when cavitation first occurs. The data are related to three types of probes, differing for diameter and surface roughness (\(Rq\) parameter). The blue vertical line represents the cavitation amplitude predicted by \citet{Pan2017} for the 3 mm probe, while the green line represents that for the 7 mm probe.} 
\label{Scatter_Amplitude} 
\end{figure}

\begin{figure}[h!]
\centering
\includegraphics[width=0.75\linewidth]{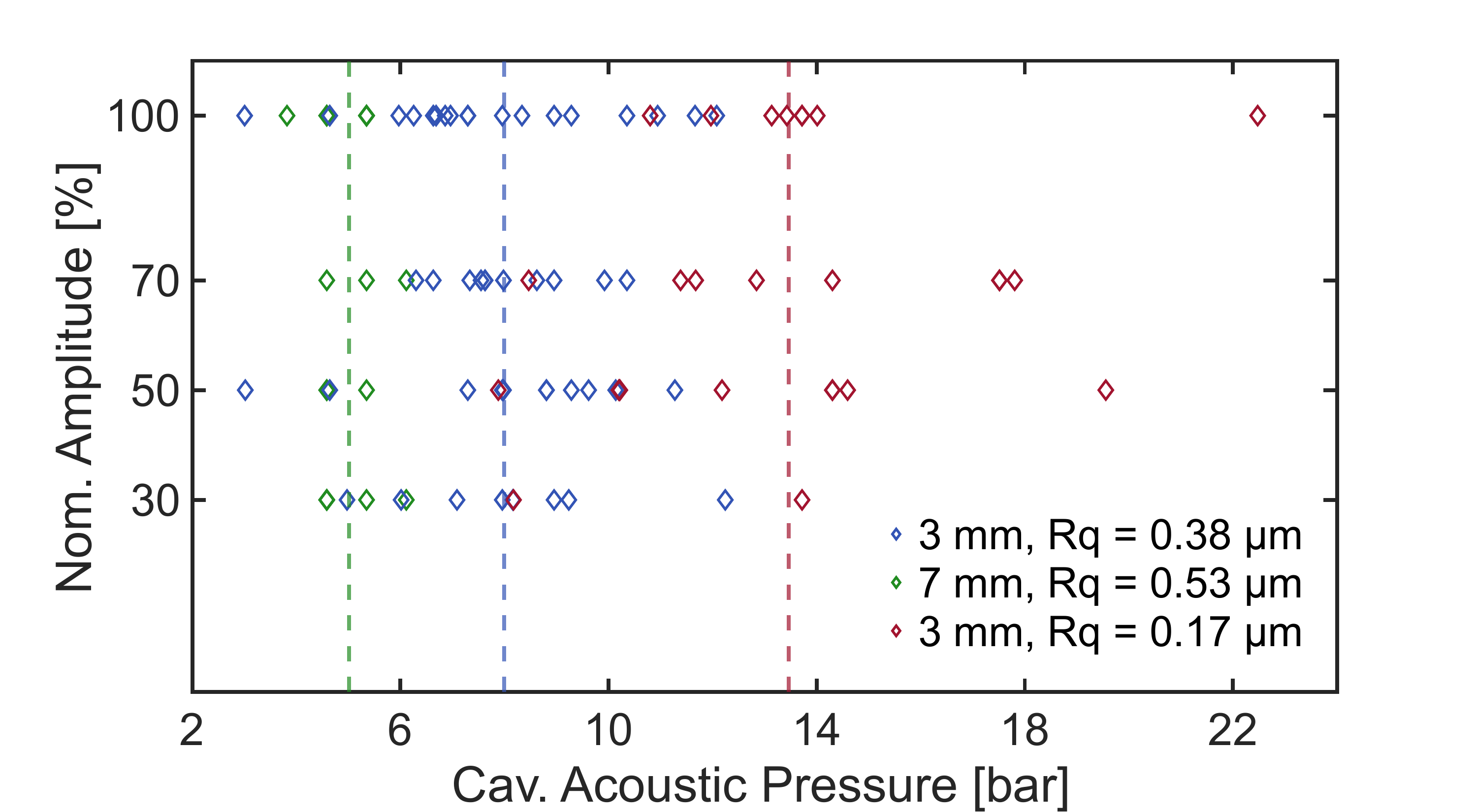}
\caption{Scatter plot, where the values of the acoustic pressure amplitude that has generated cavitation in the experimental observations (\(p_{0}|_{cav}\)). The green dashed line represents the average value of \(p_{0}|_{cav}\) (\SI{5.01}{bar}) for the 7 mm probe (\(Rq\) = 0.53). The blue line represents the average value of \(p_{0}|_{cav}\) (\SI{7.99}{bar}) for the 3 mm worn probe (\(Rq\) = 0.38), while the red one, the average (\SI{13.46}){bar} for the 3 mm smooth probe (\(Rq\) = 0.17).}
\label{Scatter_tensile} 
\end{figure}

The probe displacement was measured with a tracking procedure, implemented on Matlab \cite{MATLAB}. Starting from the sequence of frames (Fig. \ref{fig0} (A, B)), the displacement can be tracked by selecting one column of pixels and by taking the difference between adjacent pixels. The maximum value of the difference column defines the position of the probe tip at every time step. This can be used for tracking the movement over time until cavitation happens (Fig. \ref{Figure3_1} (A)). The Discrete Fourier Transform (DFT) of the data yields a probe frequency of $f_p=23.78$ kHz (close to the advertised value of the ultrasound device of 24 kHz). The red vertical line, in Fig. \ref{Figure3_1} (A), at \SI{5.2}{ms} indicates the time of cavitation inception ($t_{cav}$), e.g. \SI{20}{\micro s} in Fig. \ref{fig0} (A, B).

The consistency of the cavitation time can be confirmed by the acoustic signal of the hydrophone (Fig. \ref{Figure3_1} (B)). A sudden change happens in correspondence to the cavitation time. The hydrophone signal is symmetric to the zero during the time lapse before cavitation occurs. This indicates that the information only comes from the alternation of rarefaction and compression regions that characterize the propagation of ultrasound waves. After cavitation onset, the signal becomes asymmetric and chaotic (Fig. \ref{Figure3_1} (B)). This can be confirmed also by looking at the DFT frequency spectrum (Fig. \ref{Figure_fourier}).
The cavitation time (\(t_{cav}\)) and cavitation amplitude (\(A_{cav}\)) were then found by considering the nearest relative maxima and minima, in a neighborhood of \(t_{cav}\) (red line in Fig. \ref{Figure3_1}), and by averaging their absolute values. The procedure for finding \(A_{cav}\) has been repeated for all experimental observations: for every one of the seven configurations, the nominal amplitude of the device (\(A_{nom}\)) has been varied four times (30\%, 50\%, 70\%, and 100 \% ). The experimental run has been repeated at least twice for every combination of parameters. 
\newline

The cavitation number introduced by Pan et al. \cite{Pan2017} (Eq. \ref{equation_0_2}) can also be applied to the ultrasound transducer problem. The acceleration (\(a\)) of a vibrating transducer can be found by knowing the displacement amplitude (\(A\)) and the frequency (\(f_p\)) by \(a=A (2 \pi f_p)^2\).  Assuming the length scale factor (\(h\) in Eq. \ref{equation_0_2}) equal to the diameter of the probe, the Pan et al. number predicts a cavitation amplitude (\(A_{cav}=\frac{p_r-p_v}{\rho h (2 \pi f)^2}\)) of \SI{1.47}{\micro m} for the 3 mm probe and \SI{0.63}{\micro m} for the 7 mm probe (respectively blue and green vertical lines in Fig. \ref{Scatter_Amplitude}).
These results underestimate the value at which cavitation happens represented in the scatter plot in Fig. \ref{Scatter_Amplitude}. Indeed, among the experiments with the 3 mm worn probe (\(Rq\) = \SI{0.38}{\micro m}), the minimum and the maximum value of \(A_{cav}\) were, respectively, \SI{9.05}{\micro m} and \SI{36.71}{\micro m}. Instead, for the 7 mm probe (\(Rq\) = \SI{0.53}{\micro m}), the minimum and the maximum were \SI{4.97}{\micro m} and \SI{7.95}{\micro m}. These results highlight the necessity of introducing a more universal cavitation number to describe the acoustic system in a wide range of frequencies.

\section{Application of the Analytical Model and Results}
Starting from the vibration frequency of the probe (\(f\)) and the displacement amplitude (\(A\)), the acoustic pressure field generated by a flat acoustic transducer can be analytically estimated by solving the equation introduced by the Rayleigh Integral \citep{rayleigh1896theory, sherman, kinsler2000fundamentals}. The Rayleigh formulation is based on the linear form of the Navier-Stokes Equations, where the advection term (non-linear) is neglected. This hypothesis is generally reasonable since the velocity magnitude of the fluid particles of a sound wave is much smaller than the acceleration term \citep{Pan2017,antkowiak2007short}. The gravitational and viscous terms can be also neglected in the Momentum Equations. Furthermore, the acoustic analytical solution is based on the no-slip boundary condition, enforced between the flat surface of the transducer and the adjacent liquid medium \citep{rayleigh1896theory, sherman, kinsler2000fundamentals}. 

The cylindrical probe (Fig. \ref{Figure_end}), used for the present experimental campaign, can be modeled by considering its analogy with the acoustic radiation problem from a Plane Circular Piston \citep{rayleigh1896theory, sherman, kinsler2000fundamentals}. Assuming uniform velocity across the entire emitting flat surface, the Rayleigh integral can be analytically solved for the most relevant points of the acoustic field. An analytical solution can be obtained for the far-field and the near-field, along the axis of the probe \citep{sherman, kinsler2000fundamentals}. 

Due to the axial symmetry of the circular transducer configuration, the point of the domain with the maximum absolute value of acoustic pressure (\(p\)) lies in the near-field, on the axis of the probe \citep{sherman, kinsler2000fundamentals}. For this reason, the solution of the near-field is crucial for the investigation of cavitation phenomena \cite{sherman}. 
Assuming a cylindrical coordinate system, with radial coordinate \(r\) and axial coordinate \(x\) (Fig. \ref{Figure_end}), the complex acoustic pressure in the near-field, on the axis of the probe, \(p(r=0,x,t)\), can be found by direct integration of the Rayleigh integral \citep{sherman, kinsler2000fundamentals}: 
\begin{equation}
\label{equazione_1}
p(r=0,x,t)= -\rho c v_0 \left[e^{-i k \sqrt{(d^2+x^2)}}-e^{-i k x}\right]e^{i \omega t},
\end{equation}
where \(i\) is the imaginary unit, \(t\) is the time variable, \(\rho\) and \(c\) are respectively density and speed of sound of the liquid medium, \(v_0\) is the velocity amplitude of the probe vibration, and \(d\) is the radius of the cylindrical probe. \(v_0\) can be expressed as \(v_0=A\omega\), where \(\omega\) is the angular frequency: \(\omega=2 \pi f\). \(k\) is the wave number, which is equal to \(k=(2\pi)/\lambda\), where \(\lambda\) is the wavelength of the acoustic wave (\(\lambda=c/f\)). 
The acoustic pressure amplitude (\(p_0\)) can then be computed by considering the magnitude of the complex pressure in Eq. \ref{equazione_1}:
\begin{equation}
\label{equazione_2}
p_0= 2 \rho c v_0 \left|\sin \left[\frac{1}{2}k\left(\sqrt{(d^2+x^2)}-x\right)\right]\right|.
\end{equation}

For a circular transducer, the maximum pressure amplitude's location and intensity depend on the value of \(kd\) \citep{sherman}. If \(kd>\pi\), the maximum pressure amplitude is not on the piston surface (\(x=0\)) and it varies along the axis between zero and a maximum value of \(2 \rho c v_0\). On the contrary, if \(kd\leq\pi\), the maximum pressure amplitude occurs on the piston's surface, and the pressure amplitude value decreases monotonically along the axis \citep{sherman}. The latter is the case for the 24 kHz transducer adopted in this study (\(kd=0.15\) for the 3 mm probe, while \(kd=0.36\) for the 7 mm probe). The acoustic pressure amplitude on the surface of the transducer can be found by setting \(x=0\) in the previous Eq. \ref{equazione_2}:
\begin{equation}
p_0(x=0)= 2 \rho c v_0 \left|\sin\left(\frac{kd}{2}\right)\right|.
\label{equazione_3}
\end{equation}
If no impurities nor bubbles are present in the domain (nucleation sites), the maximum acoustic pressure defines the location of nucleation of the first cavity. This is the case for the high-speed sequences in Fig. \ref{fig0} (C, D), where the nucleation of the first cavity is attached to the surface of the tip on the axis of the probe.

Eq. \ref{equazione_3} can be used for estimating the acoustic pressure amplitude (\(p_{0}|_{cav}\)) that has generated cavitation in the experimental observations (Fig. \ref{Scatter_tensile}). Assuming that the adjacent liquid has the same velocity as the transducer surface (no-slip condition), the velocity amplitude (\(v_0\)) and the wave number (\(k\)) can be computed by using the frequency ($f_p$) and the displacement amplitude (\(A_{cav}\)), resulting from the tracking procedure. 
\newline 

\begin{figure}[h!]
\centering
\includegraphics[width=0.75\linewidth]{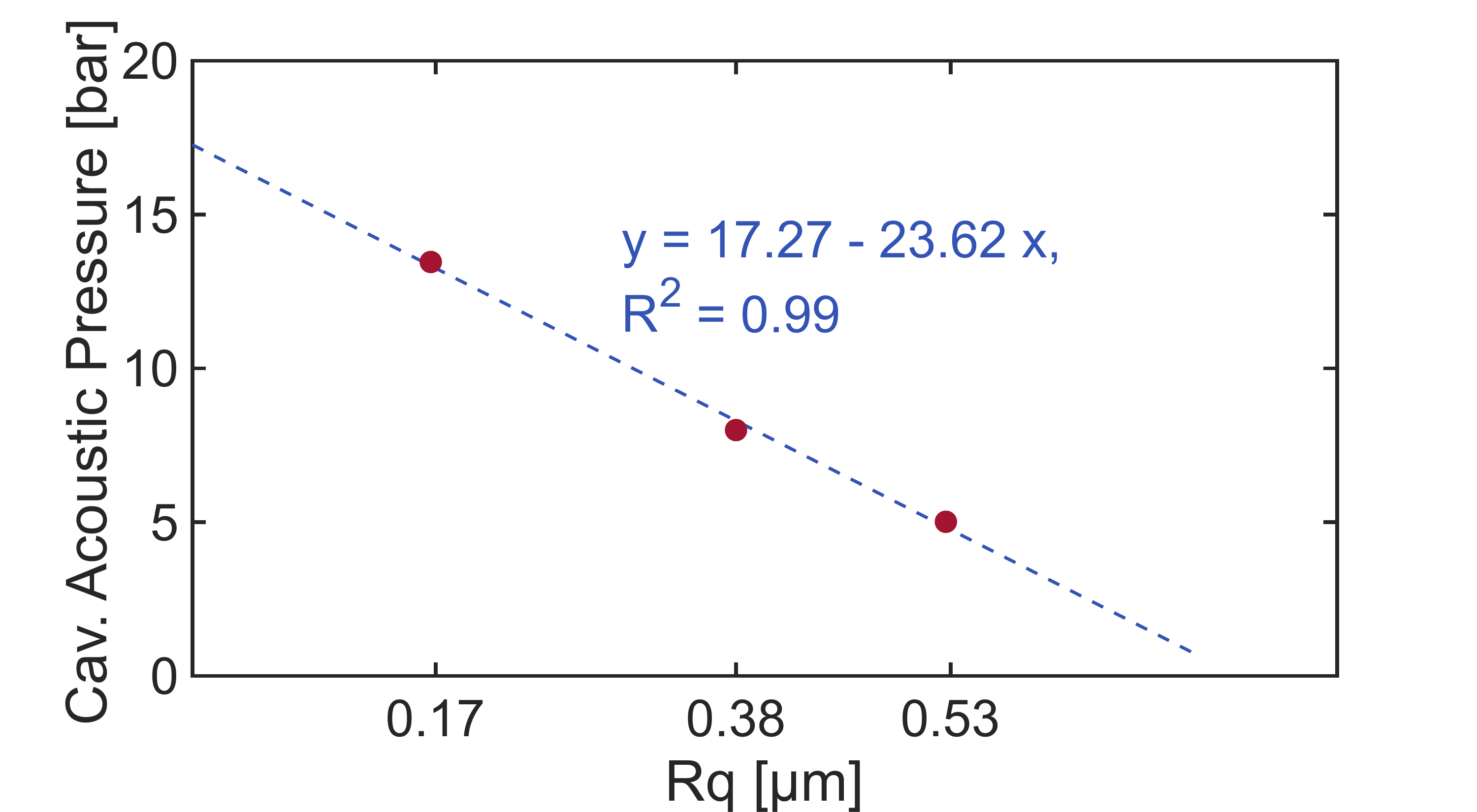}
\caption{Chart of the average value of \(p_{0}|_{cav}\) as a function of the roughness parameter (\(Rq\), Stand. JIS2001) for the three ultrasound probes tested. A linear trend (coefficient of determination of 0.99) emerges from the limited data available.}
\label{Scatter_tensile_av} 
\end{figure}

In the experimental configuration with the 7 mm probe (\(Rq\) = \SI{0.53}{\micro m}), the minimum and the maximum value of \(p_{0}|_{cav}\) are respectively \SI{3.82}{bar} and \SI{6.11}{bar}. For the 3 mm worn probe (\(Rq\) = \SI{0.38}{\micro m}), the minimum and the maximum value of \(p_{0}|_{cav}\) are \SI{3.05}{bar} and \SI{12.25}{bar}. Instead, between the experimental configuration with the 3 mm smooth probe (\(Rq\) = \SI{0.17}{\micro m}), the minimum and the maximum value of tensile strength are \SI{7.88}{bar} and \SI{22.48}{bar} (Fig. \ref{Scatter_tensile}).
The average value of \(p_{0}|_{cav}\) for the 7 mm probe (\(Rq\) = \SI{0.53}{\micro m}) is \SI{5.01}{bar}. Regarding the 3 mm probe, the average value of  \(p_{0}|_{cav}\) is \SI{7.99}{bar} for the worn probe (\(Rq\) = \SI{0.38}{\micro m}), while it is \SI{13.46}{bar} for the smooth probe (\(Rq\) = \SI{0.17}{\micro m}) (Fig. \ref{Scatter_tensile}).
These results confirm that water, forced in the ultrasound range, can withstand tension stresses before changing phase.
By looking at the data, it is clear that \(p_{0}|_{cav}\) decreases as the roughness, of the emitting surface of the probe, increases. A linear trend arises in the dependency between the average value of \(p_{0}|_{cav}\) (\(p_{0}|_{cav}^{av}\)) and the roughness parameter (\(Rq\)) which characterizes the three probes (Fig. \ref{Scatter_tensile_av}). This behavior is due to the heterogeneous nucleation mechanism (Fig. \ref{fig0}), where the surface roughness and the micro-cavities, present on the nucleation surface, reduce the energy barrier required for cavitation onset \citep{Brennen1995}. Following the formalism adopted in the present work, the value of \(p_{0}|_{cav}^{av}\) can be considered equal to the tensile strength of the liquid medium (\(p_{R}\)) in that particular condition.  

The values of \(p_{0}|_{cav}\) have not shown any clear trend in the range of DO tested (\textit{Methodology}). The range of results, obtained for the seven experimental configurations, is represented by the cyan-colored bar in Fig. \ref{figure_1} and is in good agreement with similar measurements performed by other authors \citep{esche1952untersuchung,blake1949onset,strasberg1959onset}.
\section{Definition of a new Cavitation Number}
The agreement of the results, with the experimental data (Fig. \ref{figure_1}), could suggest the method's validity. However, the previous results are based on a strong assumption: cavitation only depends on the near-field, and the reflection of the acoustic waves, on the wall of the container, can be neglected. The validity of this assumption can be verified by estimating the acoustic pressure amplitude of a wave reflecting on the wall of the container, along the axis of the probe. This can be done on the second configuration (Conf. 2, \textit{Methodology}) since it should provide the strongest pressure amplitude of the reflected wave. In this configuration, the distance between the tip of the ultrasound probe and the wall is 19.66 mm, and the acoustic reflective index between the water and the material of the wall (optical glass) is R = 0.94. Since the relative position, between the emitting surface and wall, is larger than several probe diameters; the acoustic pressure can be evaluated by using the far-field approximation \citep{sherman,kinsler2000fundamentals}: 
\begin{equation}
\label{equazione_4}
p(r,\theta,t)= \frac{i}{2} \rho c k v_0 \frac{d^2}{R} \left[ \frac{2 J_1(k d \sin{\theta})}{k d \sin{\theta}} \right] e^{i(\omega t-kR)},
\end{equation}
where \(R\) and \(\theta\) are respectively the radial and the polar coordinates of a spherical reference system with the origin on the axis of the probe, on the emitting surface (Fig. \ref{Figure_end}). The azimuthal angle does not appear in Eq. \ref{equazione_4} because of the axial symmetry (Fig. \ref{Figure_end}). The term \(J_1\) in Eq. \ref{equazione_4} is the First-Order Bessel Function \citep{sherman, kinsler2000fundamentals}. Assuming a vibration amplitude of \SI{20}{\micro m} for the ultrasound probe in Conf. 2, the maximum acoustic pressure predicted in the near field is \SI{6.69}{bar}, while the pressure amplitude of the wave reflected on the wall, along the axis of the probe, is estimated to be \SI{0.23}{bar} (\(1/R\) trend with the radial distance (Eq. \ref{equazione_4})). This clearly demonstrates that the influence of the reflected acoustic wave can be neglected in the estimation of the pressure threshold for acoustic cavitation, as long as the wall of the container is placed at a sufficiently far distance from the emitting surface (far-field). A direct validation, of the analytical model (Eq. \ref{equazione_2}), adopted for the estimation of the acoustic pressure, can be obtained by comparison with the pressure measurements done with the hydrophone. In every experimental configuration, the hydrophone was placed in a certain point of the domain, in the far-field, where the acoustic pressure amplitude can be estimated, in polar coordinates, by using Eq. \ref{equazione_4}. The comparison between the analytical results and the hydrophone data is visible in Tab. \ref{tab1}. These are related to different nominal amplitudes and instants of time for Conf. 3 (\textit{Methodology}).
\begin{table}[t!]
\centering
\begin{adjustbox}{width=\textwidth}
\begin{tabular}{lrrrrrr}
 \(A_{nom}\) [\%] & Time [ms] & Hydrophone Voltage [V] & Hydrophone Pressure [bar] & \(A\) [m] & \(p_0\) [bar] & Relative Error [\%]\\
\midrule
30 & 3 & 0.11 & 0.027 & 7.73E-06 & 0.026 & 5.36\\
30 &  3.5 & 0.155 & 0.038 &9.02E-06 & 0.030 &21.65\\
30 &  4 & 0.3 & 0.075 & 1.59E-05 & 0.054 & 28.62\\
50 & 3 & 0.15 & 0.037 & 9.70E-06 & 0.033 & 12.88\\
50 & 3.5 & 0.215 & 0.053 & 1.03E-05 & 0.035 & 35.17\\
50 & 4 & 0.415 & 0.103 & 1.29E-05 & 0.044 & 58.02\\
100 & 3 & 0.16 & 0.040 & 9.70E-06 & 0.033 & 18.33\\
100 & 3.5 & 0.27 & 0.067 & 1.10E-05 & 0.037 & 45.15\\
100 & 4 & 0.365 & 0.091 & 1.23E-05 & 0.041 & 54.65\\
\bottomrule
\end{tabular}
\end{adjustbox}
\caption{Comparison between the acoustic pressure amplitude (\(p_0\)) estimated with the analytical model and the measurement done with the hydrophone (Hydrophone Pressure). The data are related to different nominal amplitude and time instants for Conf. 3.}
\label{tab1}
\end{table}

The data shows a good agreement between the value of the acoustic pressure amplitude predicted by the model and those measured by using the hydrophone. Even the largest deviation (58.02 \% in Tab.\ref{tab1}) can be justified by taking into account all the non-idealities present in the experimental setup: the limited resolution used in the tracking procedure for the estimation of the vibration amplitude, the uncertainty related to the position of the hydrophone with the respect of the ultrasound probe, and the influence of the acoustic wave reflected on the walls of the container. The increase of the relative error, as time progresses, could be due to a larger acoustic pressure of the reflected waves since the vibration amplitude of the probe also increases with time.
\newline

Considering the experimental results obtained by the proposed method, a new cavitation number \citep{viciconte2023unifying} can be defined. This is based on the definition of tensile strength and the estimation of the acoustic pressure with the analytical model (Eq. \ref{equazione_3}) :

\begin{equation}
\label{equazione_5}
Ca = \frac{p_{R}(f,T,Rq)+\rho g h-(p_{atm}-p_r)}{ 2 \rho c A \omega \left|\sin(\frac{kd}{2})\right|},
\end{equation}
where \(\rho g h\) is the hydrostatic pressure, \(p_r\) is the reference pressure, and \(p_{atm}\) is the atmospheric one. The tensile strength of the liquid medium (\(p_R(f,T,Rq)\)) represents the cavitation threshold, replacing the vapor pressure traditionally used in the context of hydrodynamic cavitation. The formulation of the denominator in Eq. \ref{equazione_5} is only valid when using a circular transducer where \(kd\leq\pi\) \citep{kinsler2000fundamentals, sherman}. In the case of a transducer with \(kd>\pi\), the number takes the form of:
\begin{equation}
\label{equazione_5_2}
Ca = \frac{p_{R}(f,T,Rq)+\rho g h-(p_{atm}-p_r)}{ 2 \rho c A \omega}.
\end{equation}
Starting from the theoretical arguments discussed, it arises that \(p_R\), expressed in a general form, is a function of the frequency of the acoustic wave (\(f\)) and the temperature of the liquid medium (\(T\)). This dependency has not been directly investigated in the present work, since all the experiments have been run at fixed frequency and temperature of the liquid. Furthermore, \(p_R\) is also a function of the roughness parameter \(Rq\). Indeed, the experimental results (Fig. \ref{Scatter_tensile}) demonstrate that the cavitation threshold, for heterogeneous nucleation, depends on the surface roughness. This dependency leads to a reduction of the acoustic cavitation threshold compared to the case of homogeneous nucleation. Considering that Eq. \ref{equazione_5} and \ref{equazione_5_2} are related to a plane circular piston, we can further generalize the equation generalized the cavitation number:

\begin{equation}
\label{equazione_6}
Ca = \frac{p_{R}(f,T,Rq)+\rho g h-(p_{atm}-p_r)}{ C_g \rho c A_0 \omega}.
\end{equation}
The term \( \rho c A_0 \omega \) represents the acoustic pressure amplitude of a plane wave, solution of the one-dimensional acoustic wave equation \citep{lurton, kinsler2000fundamentals}:
\begin{equation}
\label{equazione_7}
\frac{\partial^2 p}{\partial t^2} = c^2 \frac{\partial^2 p}{\partial x^2},
\end{equation}
while \( C_g \) (Eq. \ref{equazione_7}) is a non-dimensional coefficient which is a function of the transducer geometry (for a circular piston \( C_g=2\left|\sin\left[\frac{kd}{2}\right]\right|\)). \( C_g\) acts as a corrective factor, to make the denominator equal to the maximum acoustic pressure amplitude generated by a particular transducer geometry. 
It is important to clarify the above definition of the cavitation number (Eq. \ref{equazione_6}) is based on the assumption that the medium has been characterized and the tensile strength is known for a given type of liquid, under certain working conditions (\(f,T,Rq\)). However, considering the lack of experimental data available in the literature \citep{yakupov2023application}, this assumption is unrealistic for most applications. For these reasons, the vapor pressure remains a reasonable cavitation threshold, especially for engineering applications where a safety coefficient is required or for working liquids that are particularly dirty or with a high presence of impurities \citep{plesset1969tensile}.

\section{Acoustic analogy to describe impulsive cavitation}
The same approach proposed can be applied to study phenomena where an impulsive motion causes cavitation. We used the set of mallet-tube
experiments, proposed by \citet{Pan2017}, to demonstrate the validity of the acoustic analogy for impulsive phenomena. Pan et al. investigated the cavitation activity that occurs when a liquid-filled bottle is struck, from the top, by a mallet. Under the action of the impulsive force given by the mallet, the rapid acceleration of the bottle creates a rarefaction region in the liquid medium. This induces the formation and collapse of tiny bubbles at the bottom of the bottle \citep{Pan2017}. The classical velocity-based cavitation number Eq.~\ref{equation_0_1}, cannot represent cavitation in liquids that undergo an impulsive acceleration. For this reason, the authors introduced Eq. \ref{equation_0_2}. The experimental data, considered in this section, are those collected at Utah State University and Brigham Young University (USU/BYU) \citep{daily2013fluid, daily2014catastrophic,Pan2017}. For the experimental setup, the USU/BYU group used cylindrical transparent tubes, having an inner diameter ($D$) of \SI{55}{\milli\meter}. The tube was partially filled with distilled water and the internal reference pressure (\(p_r\)) was kept, through a pressure tap, to a value of \SI{86.9}{kPa}. The water properties considered by the authors can be found in \citep{Pan2017}. 
The height of the liquid (\emph{h}), within the tube, was varied in a range from \SI{12.5}{\milli\meter} to \SI{74.5}{\milli\meter}. 

The tube was accelerated by striking the top with a rubber mallet. For every experiment, the authors monitored the presence of cavitation bubbles, using a high-speed camera, and the acceleration of the tube over time, by placing an accelerometer on the tube bottom. The sampling interval of the accelerometer was \SI{0.2}{\milli\second}. Additional details on the experimental setup used by the USU/BYU group can be found in \cite{Pan2017}. While Pan et al. just considered the maximum acceleration (\(a_{max}\)) measured by the hydrophone, to reinterpret the phenomenon, through an acoustic analogy, we must take into account the entire acceleration curve over time. By doing this, the acceleration function can be expressed as a summation of sine and cosine, and the characteristic frequencies, associated with the impulsive motion of the tube, can be extracted. 

\begin{figure}[h!]
\centering
\includegraphics[width=0.75\linewidth]{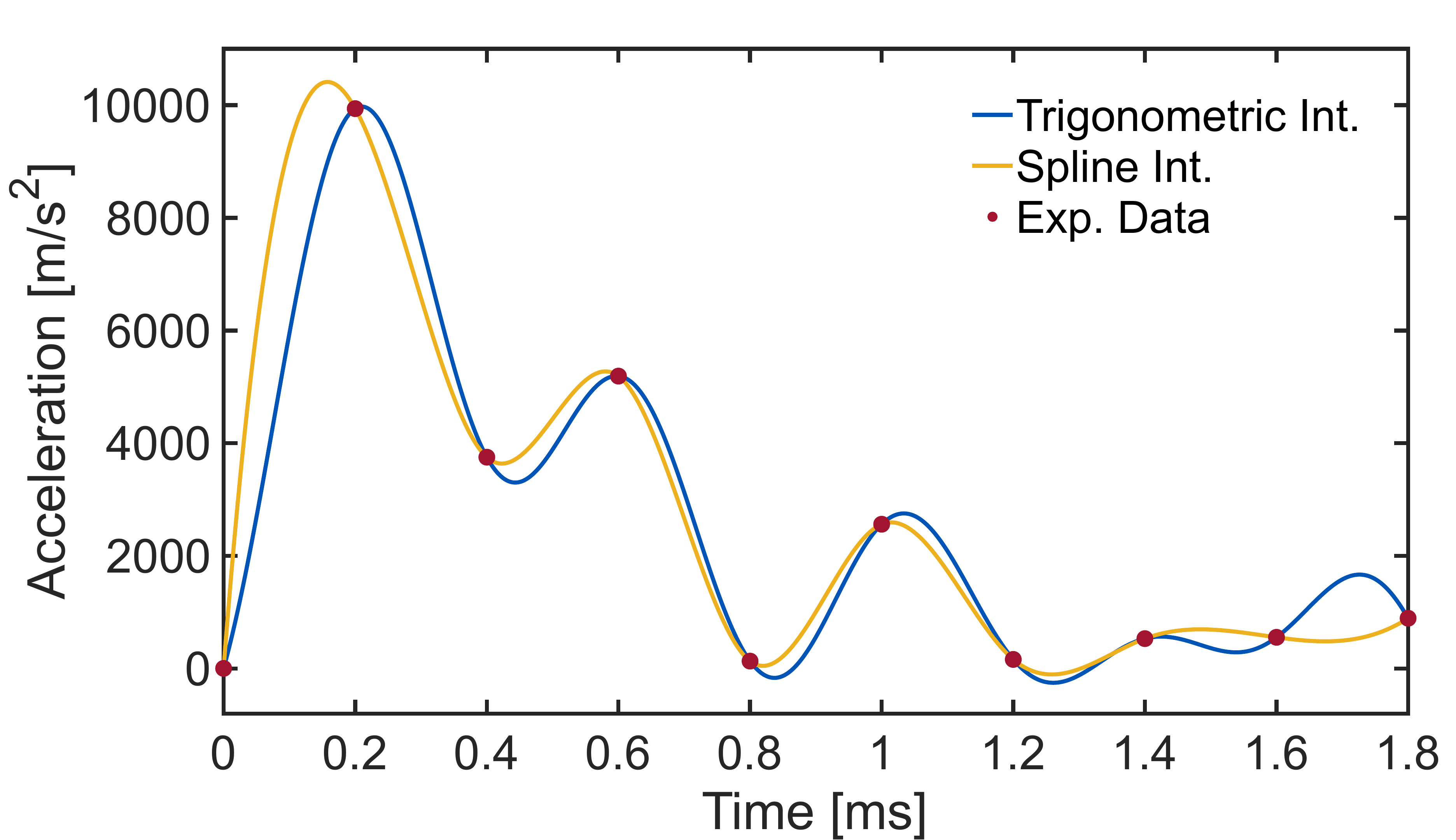}
    \caption{Experimental data, spline, and trigonometric interpolation of the acceleration of the cavitation tube.} 
\label{pic_acc2} 
\end{figure}

For every observation, the global acceleration data set was "cleaned" by considering just a temporal window of \SI{1.8}{\milli\second} in a neighborhood of the main acceleration impulse (Fig. \ref{pic_acc2}). This is equivalent to taking just ten experimental data (red dots in Fig. \ref{pic_acc2}), but it is enough to represent the relevant acceleration impulse induced by the mallet impact on the tube. The acceleration values (Fig. \ref{pic_acc2}) are positive since the coordinate of the reference system of the accelerometer coincides with the axis of the tube and faces towards the outside \citep{Pan2017}.

Considering the characteristic frequencies, associated with the impulsive acceleration of the tube, the data can be interpolated by a summation of trigonometric functions \citep{sauer2011numerical}. The acceleration is therefore defined as:
\begin{equation}
\label{equazione_8}
\begin{split}
a (t) & = \frac{1}{\sqrt{n}} 
      [\Re(y_{1})+\cos (\pi  m  t S_f)\Re(y_{n/2+1})]+ \\
      &+ \frac{2}{\sqrt{n}} \sum _{j=1}^{\frac{m}{2}-1} [\cos (2 \pi  j t S_f)\Re(y_{j+1})-\sin (2 \pi  j t S_f)\Im(y_{j+1})],
\end{split}
\end{equation}
where \( j\) is the summation index, \( n\) is the number of data points, \( m\) is the order of the trigonometric function, \(y_j\) is the j-th value of the DFT, and \(t\) is the time variable. \(\Re(y_j)\) represents the real part of the j-th DFT value and \(\Im(y_j)\) represents the imaginary part. The trigonometric interpolation of the discrete data is obtained for \( m=n\). \(S_f\) is a scale factor which is a function of the sampling interval of the data, \(\Delta t_d\) (equal to 0.2 ms in the current case), through the following relation: 
\begin{equation}
\label{eq_26}
S_f=\frac{0.1}{\Delta t_d}.
\end{equation}
The blue curve in Fig. \ref{pic_acc2} represents the trigonometric interpolation of the experimental data (red dots in Fig. \ref{pic_acc2}).
The interpolation provided by Eq. \ref{equazione_8} \citep{sauer2011numerical} is similar to the Cubic Spline Interpolation \citep{MATLAB} (orange curve in Fig. \ref{pic_acc2}), and it seems to give a good representation of the experimental data (Fig. \ref{pic_acc2}). 

\begin{figure}[h!]
\centering
\includegraphics[width=0.75\linewidth]{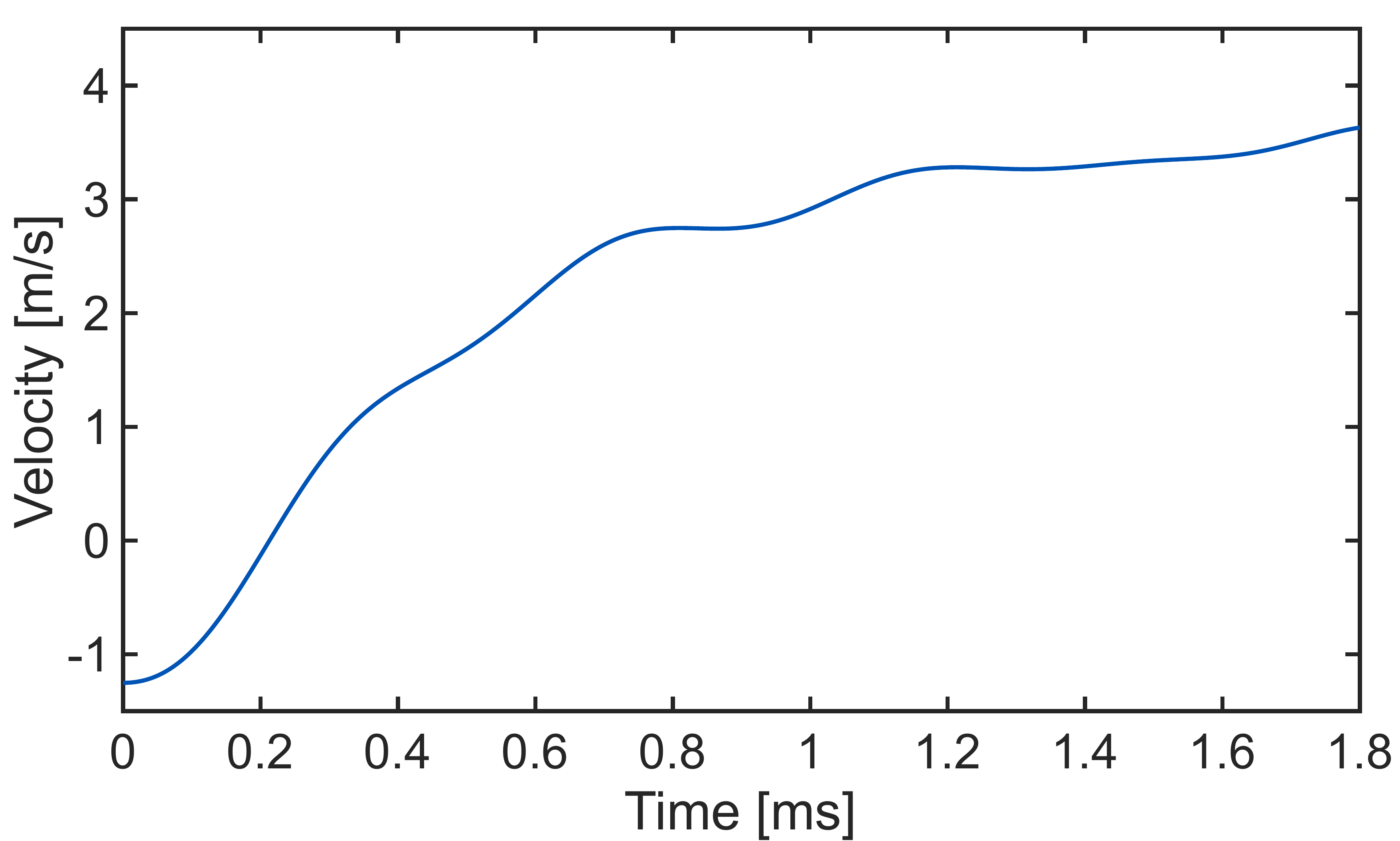}
    \caption{Trigonometric interpolation of the velocity function. The function is derived by integration of the trigonometric interpolation \citep{sauer2011numerical} of the acceleration experimental data \citep{Pan2017}.} 
\label{pic_vel} 
\end{figure}

It is important to specify that the maximum frequency, described by the trigonometric interpolation (Eq. \ref{equazione_8}), is \SI{2.5}{kHz} (\((\pi  m S_f)/(2 \pi) \)), while the sampling frequency of the experimental measurement is \SI{5}{kHz} (\(1/(\Delta t_d)\)). This is at the limit of validity of the Nyquist–Shannon Sampling Theorem \citep{gruber2013proofs}. However, to avoid any undersampling problem, having a higher sampling frequency would be better. The velocity function can be found by integrating Eq. \ref{equazione_8} in time and by setting the integration constant equal to zero. All the terms arise as sine or cosine functions, except one which is linear in time. Since the acceleration impulse of the tube lasts a short interval of time (1.8 ms), it is possible to approximate the linear term as a sine one (Taylor expansion \citep{pourahmadi1984taylor}): \((1/\sqrt{n}) \Re(y_{1}) t=(1/\sqrt{n})\Re(y_{1}) \sin(t)\). Therefore, the velocity function can be entirely expressed as a summation of sine and cosine terms (Fig. \ref{pic_vel}): 

\begin{equation}
\label{equazione_8_2}
\begin{split}
v (t) & = \frac{1}{\sqrt{n}} 
      [\Re(y_{1})\sin t+ \frac{\Re(y_{n/2+1})}{\pi m S_f}\sin(\pi  m  t S_f)]+ \\
      &+ \frac{2}{\sqrt{n}} \sum _{j=1}^{\frac{m}{2}-1} [\frac{\Re(y_{j+1})}{2 \pi  j S_f}\sin (2 \pi  j t S_f)+\frac{\Im(y_{j+1})}{2 \pi  j S_f}\cos (2 \pi  j t S_f)].
\end{split}
\end{equation}
A characteristic frequency can be associated with every term of the function. This will simplify the solution of the wave equation for estimating the acoustic pressure within the tube.
To describe this phenomenon, the mallet-tube problem can be modeled considering the plane circular transducer framework (Eq. \ref{equazione_1}) \citep{sherman, kinsler2000fundamentals}. The acoustic analogy considers the tube's bottom as a vibrating surface of the transducer and the no-slip condition between the tube and the liquid medium. Under this analogy, the trigonometric interpolation of the velocity of the tube (Eq. \ref{equazione_8_2}) can be used as a boundary condition for the solution of the acoustic pressure (Eq. \ref{equazione_3}), derived from the Rayleigh's integral \citep{rayleigh1896theory}. The solution can be found separately for every trigonometric term of the velocity function (Eq. \ref{equazione_8_2}) and the global solution can be found by applying the Superposition Principle.

\begin{figure}[h!]
\centering
\includegraphics[width=0.75\linewidth]{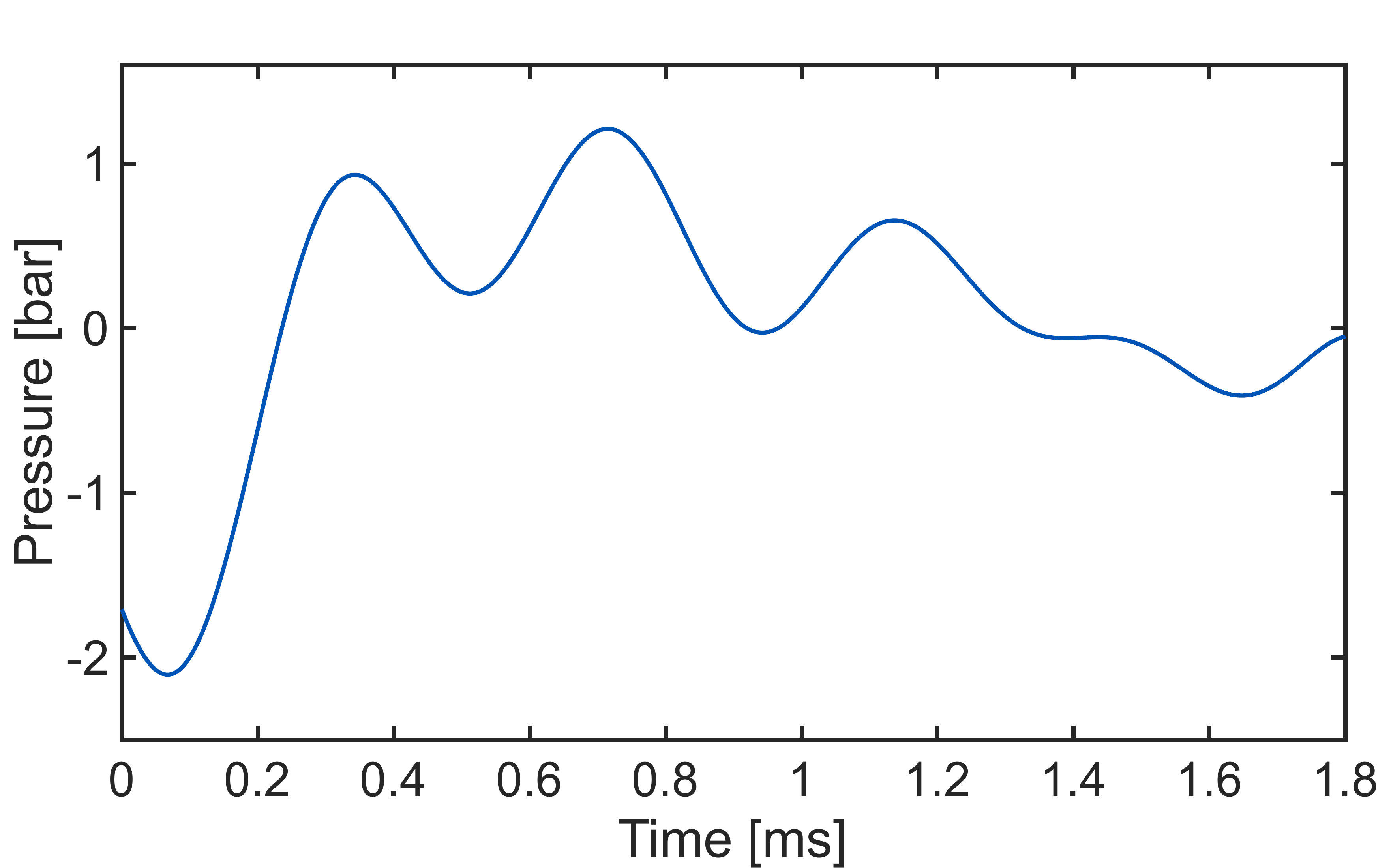}
     \caption{Acoustic pressure at the center of the bottom of the tube. This solution can be found by applying the superposition principle on every trigonometric term of the velocity function (Fig. \ref{pic_vel}).} 
\label{pic_pressure} 
\end{figure}

Since the maximum characteristic frequency, that describes the velocity function (Eq. \ref{equazione_8_2}), is equal to \SI{2.5}{kHz} (\(kd=0.30\)); the maximum pressure amplitude occurs at the center of the tube bottom. Therefore, the acoustic pressure can be derived by using Eq. \ref{equazione_3} and by considering the velocity amplitude \(v_{0i}\) associated with every term of the trigonometric velocity function (Eq. \ref{equazione_8_2}). In a generic form, the acoustic pressure (\(p(r=0,x=0,t)\)) is a summation of every $i$-th contribution, coming from the velocity equation:

\begin{equation}
\label{eq_3sgs1}
\begin{split}
p(r=0,x=0,t)=\sum _{i=1}^{n}\left[  2 \rho c v_{0i} \left|\sin\left(\frac{k_id}{2}\right)\right|\sin(\omega_i t+\phi_i) \right].
\end{split}
\end{equation}
For the case represented in Fig. \ref{pic_acc2}, the minimum of the acoustic pressure, at the center of the tube's bottom, is equal to \(p_{min}= - 2.10\) bar (Fig. \ref{pic_pressure}). 

The negative value means that the liquid locally experiences a tensile deformation. The minimum acoustic pressure was computed for all the experiments carried out by the USU/BYU group \citep{Pan2017, daily2014catastrophic, daily2013fluid}. The tensile stress (\(p_{T}\)) is considered here instead of the acoustic pressure. \(p_{T}\) can be computed by considering the presence of the hydrostatic contribution (\(p_{hyd}=\rho g h\)) and the reference pressure (\(p_{r}\)):

\begin{equation}
\label{eq_31}
\begin{split}
p_{T}=\lvert p_{min}+\rho g h-(p_{atm}-pr)\rvert.
\end{split}
\end{equation}

\begin{figure}[h!]
    \centering
\includegraphics[width=0.70\linewidth]{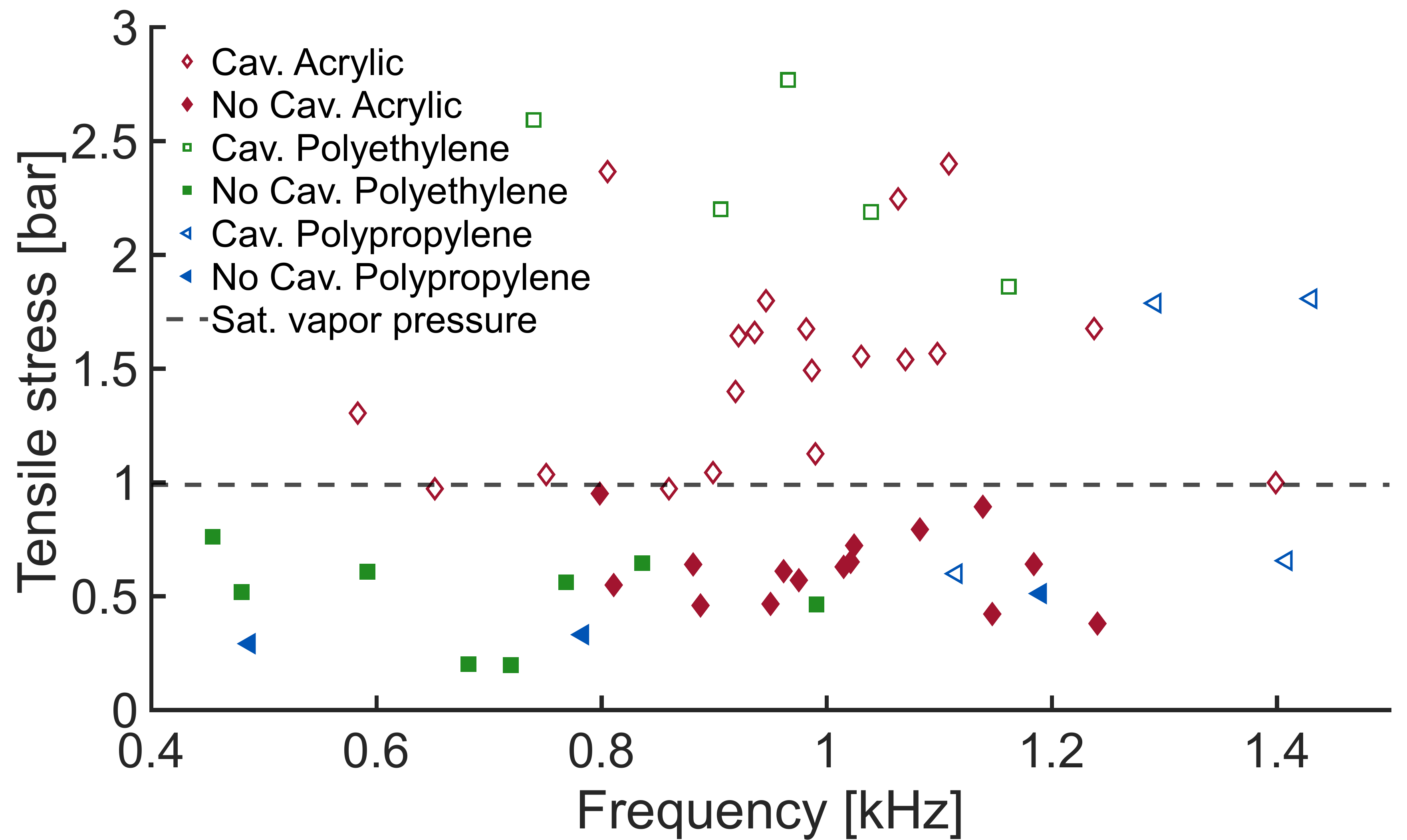}
    \caption{Scatter plot derived from the application of the acoustic analogy on the mallet-tube experimental data, collected by the USU/BYU group \citep{daily2013fluid,daily2014catastrophic,Pan2017}. The open markers denote the presence of cavitation in the experimental observation, while the filled markers indicate the absence of cavitation. The different marker colors indicate tube material: acrylic (red), polyethylene terephthalate (green), and polypropylene (blue). These experiments were conducted by the authors \citep{Pan2017} to investigate the influence of the tube material on cavitation.} 
\label{pic_sca_mal} 
\end{figure}

It is important to note that, in the above definition, the tensile stress is relative to the standard atmospheric pressure, following the convention used by \citet{urick} for the experimental chart in Fig. \ref{figure_1}. Since the USU/BYU group monitored the cavitation inception \citep{Pan2017, daily2014catastrophic, daily2013fluid}, for every experimental observation, it is possible to know whether the maximum acoustic tensile stress was able to generate or not cavitation at the bottom of the tube. This is synthesized in the scatter plot in Fig. \ref{pic_sca_mal}, where the maximum tensile stress is plotted against the weighted average frequency of the pressure wave produced by the impulsive motion of the mallet. The open markers denote the presence of cavitation in the experimental observation, while the filled markers indicate the absence of cavitation activity. 
Since in the formalism adopted, the tensile stress is defined with respect to the standard atmospheric pressure (Eq. \ref{eq_31}), the absolute vapor pressure (\SI{2338}{Pa} at \SI{20}{^\circ C}) corresponds to a tensile stress of \SI{0.99}{bar} (dashed horizontal line in Fig. \ref{pic_sca_mal}). Observing the chart in Fig. \ref{pic_sca_mal} it is clear that the vapor pressure can be considered as a reasonable cavitation threshold, since it separates the cavitation region, defined by the open markers, from the region without cavitation, defined by the filled markers.
The threshold value of  \SI{0.99}{bar} can be compared with the experimental chart (red marker in Fig. \ref{figure_1}), after being associated with the characteristic frequency of the phenomenon (from \SI{0.48}{kHz} to \SI{1.43}{kHz}, Fig. \ref{pic_sca_mal}). This is in agreement with the average value of \SI{1}{atm} reported, for the same range of frequency, by \citet{esche1952untersuchung} (Fig. \ref{figure_1}). The results demonstrate that the vapor pressure of water can be taken as a cavitation threshold for problems characterized by low frequencies. This explains the success of the proposed cavitation number by \citet{Pan2017} (Eq. \ref{equation_0_2}), where the vapor pressure appears as a threshold in predicting cavitation for the mallet-tube data set.
\newline

The cavitation number, introduced before in Eq. \ref{equazione_5}, is valid when a singular frequency characterizes the acoustic system. We can also introduce a cavitation number with multiple characteristic frequencies where the summation of $n$ trigonometric functions is slightly more complicated:
\begin{equation}
\label{eq_32}
\begin{split}
Ca = \frac{p_{R}(f_{av},T,Rq)+\rho g h-(p_{atm}-pr)}{\left| min\left(\sum _{i=1}^{n}\left[  2 \rho c v_0^i \left|\sin(\frac{k_id}{2})\right|\sin(\omega_i t+\phi_i) \right]\right)\right|},
\end{split}
\end{equation}
the $i$-th elements are related to every trigonometric term of the velocity functions that describe the vibrating surface, and $\phi_i$ is the phase of the trigonometric function. 

\section{Conclusions}

We introduced a novel cavitation number derived from the observation that in acoustic cavitation phenomena, the vapor pressure is not a suitable threshold for the phase change. Instead, the proposed dimensionless number is based on the definition of the tensile strength of a liquid medium and on the acoustic pressure, derived by using the Rayleigh analytical model. The characteristic frequency of the phenomenon appears as a scaling factor in the cavitation number both in the denominator of the acoustic pressure term and in the numerator as an independent variable of the cavitation threshold. The cavitation onset was defined using a procedure based on high-speed imaging and the acoustic signal measured by a hydrophone. The two measurements observe the same instant in time that cavitation occurs. 
This methodology could be used to study the cavitation onset in other types of acoustic systems, characterized by single and/or multiple frequencies. Further, when the methodology was applied to a cavitation event induced by a single impulsive action of a mallet on a tube containing water, the acoustic analogy accurately described the phenomenon.

\section*{Methodology}

For the generation of ultrasound waves, in the liquid domain, an Hielscher device UP400S was used. This device works at a fixed nominal frequency of \SI{24}{kHz}, and it can provide a maximum power of \SI{400}{W}. A variable vibration amplitude can be set as a percentage of the maximum one. Cylindrical probes made of titanium have different diameters and can be installed on the ultrasound device. For the current study, two different cylindrical probes, having diameters of \SI{7}{mm} and \SI{3}{mm}, have been used. Furthermore, both worn and new probes were tested. An experimental setup, based on the backlighting technique and high-speed imaging, was adopted to capture the cavitation inception during the vibration of the ultrasound probe (Fig. \ref{Figure_end}). A LED light source (Godox SL-200W II), emitting white lite non-coherently, was placed on the back side of a water container and aligned with the axis of the ultrasound probe, which is immersed in the liquid water. The scene was captured using a high-speed camera (Photron Nova S16) placed in front of the container, to the opposite side of the LED light source. The camera was triggered using the signal coming from a hydrophone (Aquarian AS-1 Hydrophone, calibration factor: \SI{40}{\micro V}/Pa), and passing through an oscilloscope. All the experiments used water with a temperature of 24$^{\circ}$ C. The liquid water's dissolved oxygen content (DO) was not altered with preliminary treatments. The experiments were conducted with a level of DO saturation percentage oscillating between 64\% and 84\%. Ultrasound probes, having diameters of 3 mm and 7 mm have been used. Furthermore, three different levels of surface roughness, for the emitting surface of the probes, were tested. The roughness has been characterized using the surface roughness tester Mitutoyo Surftest SJ-400, and by considering the root mean square deviation of the roughness (\(Rq\), Stand. JIS2001). Seven different configurations have been tested during the experimental campaign. These configurations differ for type of water, geometry and material of the transparent container, immersion depth of the probe, and relative position between the probe and the hydrophone. In the first (Conf. 1) and second configuration (Conf. 2) a 3 mm cylindrical probe and a transparent container made of optical glass and having an optical path of 50 mm (Hellma) were used. The water level in the container was 39 mm for both the configurations. The experiments related to Conf. 1 were run using distilled water from a local vendor, together with an immersion depth of the probe of 16.4 mm, and a distance between the axis of the probe and the acoustic center of the hydrophone of 34.7 mm. For Conf. 2, the immersion depth was set to 19.5 mm, and the probe/hydrophone distance to 34.5 mm. In this configuration an extra pure water, deionized (Thermo Fisher), was used. In the third (Conf. 3) and fourth (Conf. 4) configurations, Milli-Q water was placed in an transparent acrylic container, having a transversal section of 200 mm\(\times\)200 mm. The water level in the container was kept at 80 mm. In Conf. 3, a 3 mm cylindrical probe was used and submerged 20 mm below the free surface of the water. The distance between the probe and the acoustic center of the hydrophone was 74.2 mm. In Conf. 4, a 7 mm probe and an immersion depth of 20 mm have been used. In this case, the distance between the probe axis and the hydrophone was set to 70.7 mm. In the last three experimental configurations, Conf. 5, 6, and 7; a container made of optical glass (Hellma, 50 mm optical path) was used, together with a 3 mm probe, submerged 20 mm below the free surface. The types of water used were: distilled water from a local vendor (Conf. 5), extra pure water, deionized (Thermo Fisher), and pure water demineralized (Thermo Fisher). In Conf. 3, a Leica zoom lens (12x) was adopted. This lens can guarantee a level of magnification of 12x, providing a resolution of \SI{1.28}{\micro m}. In the other configurations, a Mitutoyo 10X objective lens was used. This lens can guarantee a resolution of \SI{1.98}{\micro m}. Since the nominal frequency of the ultrasound device is 24 kHz, the scene was recorded using a frame rate of 200000 fps (200 kHz). The Nova S16 camera can guarantee an image size of 128x224 pixels at this frame rate. The shutter speed of the camera was set to 1/500000 s.

\bibliographystyle{unsrtnat}
\bibliography{pnas-sample}  
\end{document}